\documentclass[twoside]{article}
\usepackage{Proc_NTSE_18-arXiv}
\begin{document}
\thispagestyle{plain}
\publref{Kim}

\begin{center}
{\Large \bf \strut\boldmath
Daejeon16 $NN$ Interaction\label{ykim}
\strut}\\
\vspace{10mm}
{\large \bf\strut
Y.~Kim$^a$, I.~J.~Shin$^a$, A.~M.~Shirokov$^{b,c}$, M.~Sosonkina$^d$, P.~Maris$^e$ and J.~P.~Vary$^e$\strut}
\end{center}
\noindent{
\small $^a$\it Rare Isotope Science Project, Institute for Basic Science, Daejeon 34047, Korea} \\
{\small $^b$\it Skobeltsyn Institute of Nuclear Physics, Moscow State University, Moscow 119991, Russia} \\
{\small $^c$\it Department of Physics, Pacific National University, Khabarovsk 680035, Russia} \\
{\small $^d$\it Department of Modeling, Simulation and Visualization Engineering, Old Dominion Univer-
\hphantom{$^d$}sity, Norfolk, VA 23529, USA} \\
{\small $^e$\it Department of Physics and Astronomy, Iowa State University, Ames, IA 50011, USA}

\markboth{Y. Kim \it et al.}{Daejeon16 $NN$ interaction}

\begin{abstract}
We have developed a realistic nucleon-nucleon ($NN$) interaction, dubbed Daejeon16.
We start from a SRG (similarity renormalization group) evolved chiral N3LO interaction.
We then apply 
PETs (phase-equivalent transformations) to the SRG-evolved interaction. It turned out that the
obtained in such a way
Daejeon16 $NN$ interaction
provides a good description of various observables in light nuclei without $NNN$ forces.
In this contribution, we present our new results for some selected nuclei using the {\it ab initio} no-core shell
model (NCSM) with the Daejeon16 interaction. One of the interesting results is that the {\it ab initio} NCSM
with Daejeon16 clearly demonstrates the phenomenon of parity inversion in $^{11}$Be,  i.\:e., the ground state in
$^{11}$Be has the spin-parity $1/2^+$ in experiments contrary to the expectation from the conventional shell model.
\\[\baselineskip]
\end{abstract}

\section{Introduction}

As the advent of new rare isotope (RI) facilities such as FAIR, FRIB, HIRFL, NICA, RAON,
etc., we have much more opportunities to resolve big questions in science.
Nuclear theory for rare isotopes should be timely developed to face new precise observables
from the forthcoming RI facilities which can produce exotic nuclei near the nuclear drip
line.
Thanks to the rapid developments of high performance supercomputers, we have a good
chance to conduct a rigorous study of nuclear structures and reactions using fundamental
(or realistic) nuclear interactions based on quantum chromodynamics (QCD). Several promising {\it ab initio} methods have
been developed for nuclear structure and reactions~\cite{Pieper:2001mp,Kowalski:2003hp,Barrett:2013nh,Epelbaum:2011mdykim,Navratil:2009ut}.

The {\it ab initio} theory requires a high-quality realistic inter-nucleon interaction to provide
 predictions for binding energies, spectra and
other observables in nuclei with mass up to A $\sim 20$ and selected heavier nuclear systems around
closed shells. Therefore, it is important to
develop realistic nucleon-nucleon ($NN$) interactions with better convergence that require less computational
resources.
In Ref.~\cite{Shirokov:2016ead}, we developed a realistic $NN$ interaction, dubbed Daejeon16,
starting from a chiral N3LO interaction which is SRG (similarity renormalization group) evolved.
Then, we apply 
PETs (phase-equivalent transformations) to the SRG-evolved
interaction. It turned out that Daejeon16 provides a good description of various observables in
light nuclei without $NNN$ forces and also generates rapid convergence in {\it ab initio} calculations.

In this short write-up, after a brief description of Daejeon16, we present some recent results from
{\it ab initio} nuclear studies using the Daejeon16 $NN$ interaction, with an emphasis on the ground-state parity inversion in $^{11}$Be.

%
%

\section{Daejeon16 and applications}

\begin{figure}[t!]
\centerline{\includegraphics[width=0.7\textwidth]{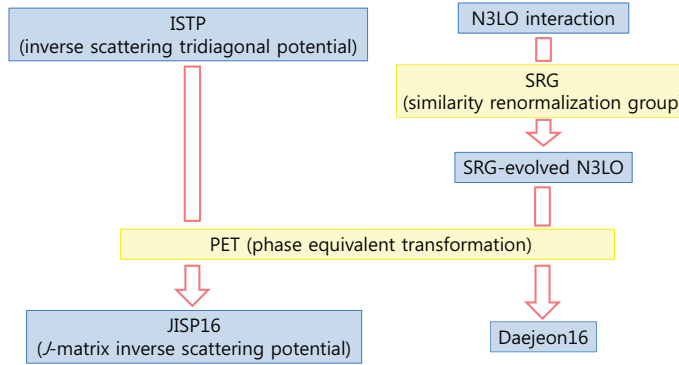}}
\vskip 1cm
\caption{A sketch of the procedure to obtain Daejeon16 compared with JISP16.}
\label{D-16}
\end{figure}

Nuclei are composed of nucleons (protons and neutrons) and their properties such as the binding energy and radius are largely governed by
the nuclear force,  i.\:e., the strong interaction. Therefore, nuclear forces are at the core of the nuclear structure studies.
Meson exchange theory has been successfully applied to obtain realistic nuclear potentials such as CD-Bonn, Argonne V18, etc.
A bit more down to the mother theory of the nuclear force,  i.\:e., QCD, nuclear interactions
from the chiral effective field theories have been developed: N2LO, N3LO, etc.
For {\it ab initio} nuclear studies, the JISP16 interaction~\cite{Shirokov:2005bk}, which is phenomenological, has been widely used.
Here, JISP stands for $J$-matrix Inverse Scattering Potential.
Recently, a new nuclear force dubbed `Daejeon16' has been developed from a N3LO $NN$ interaction.
Daejeon is a city in Korea where a next generation RI facility called RAON will be built 
and 16
  is from $^{16}$O which is the heaviest nucleus used in fitting process.
In Ref.~\cite{Shirokov:2016ead}, the authors start from Idaho N3LO $NN$ interaction and apply to it PETs which preserve 
scattering phase shifts and bound state energy of the two-nucleon system (deuteron). 
The optimal set of PET parameters is determined to describe the
binding energies of $^3$H, $^4$He, $^6$Li, $^8$He, $^{10}$B, $^{12}$C and $^{16}$O nuclei and  excitation energies of a few narrow excited states: the two lowest excited states with (J$^\pi$, T) =(3$^+$, 0)
 and (0$^+$, 1) in $^6$Li and the first excited states (1$^+$, 0) in $^{10}$B and (2$^+$, 0) in $^{12}$C.
For a sketch about the procedure to obtain JISP16 and Daejeon16, we refer to Fig.~\ref{D-16}.

\begin{figure}[b!]
\centerline{\includegraphics[width=0.7\textwidth]{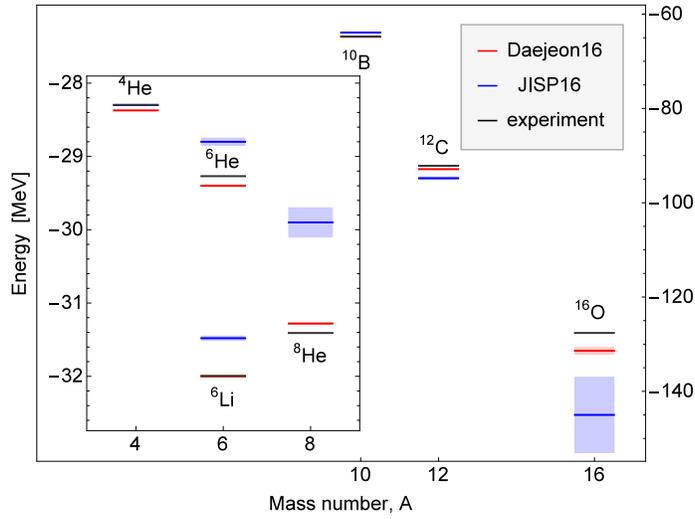}}
\caption{The ground state energies of several $p$-shell nuclei using Daejeon16 and JISP16 compared with experiment.
The calculations were performed within the NCSM 
and extrapolated to the infinite basis space using the methods of Ref.~\cite{Maris:2008ax}; the shaded areas show the uncertainties of the extrapolations.
It is noted that all shown nuclei were used to PET fitting as mentioned in the text.}
\label{comparison}
\end{figure}

It turned out that the
Daejeon16 works well for light $p$-shell nuclei compared with other established interactions such as JISP16, for instance see Fig.~\ref{comparison}.

Now, we move on to some recent results using the Daejeon16 interaction.

\subsection{Parity inversion in \boldmath$^{11}$Be}

\begin{figure}[t!]
\centerline{\includegraphics[width=0.7\textwidth]{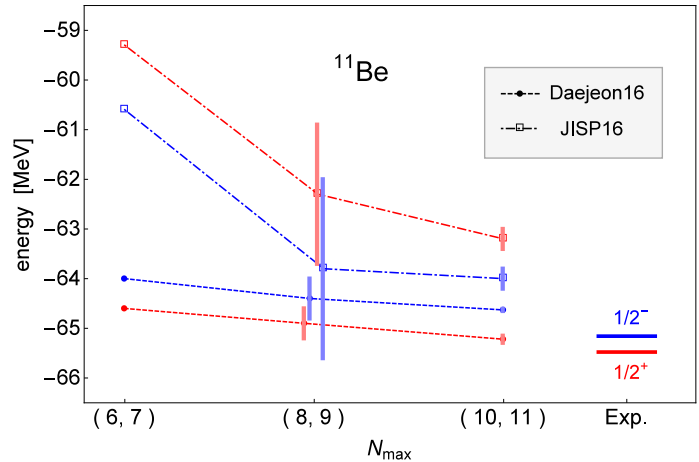}}
\caption{Energies of the ground and first-excited states of $^{11}$Be calculated within the {\it ab initio}
NCSM 
with Daejeon16 and JISP16. The values are obtained using extrapolation B~\cite{Maris:2008ax}
for each highest $N_{\rm max}$ at 
 the variational minima and the error bars are given as the differences with the
previous $N_{\rm max}$ extrapolation. Experimental values are taken from Refs.~\cite{Kelley:2012qua,AME2016:Table}.}
\label{11Be}
\end{figure}

$^{11}$Be shows an interesting feature which is opposite to the expectation from the conventional shell model.
Experimentally, the ground state of $^{11}$Be is $1/2^+$~\cite{Kelley:2012qua}, while it was expected to be a $1/2^-$ state in the conventional shell model. 
To tackle the issue of the parity inversion in $^{11}$Be, we evaluate the spectrum of $^{11}$Be using the {\it ab initio} no-core shell model (NCSM)
with the Daejeon16 interaction and extrapolate the results to the infinite basis
space using the method of Ref.~\cite{Maris:2008ax}. For the two lowest-lying states, we obtain
\begin{equation}
 1/2^+:~ -65.22(7) ~{\rm MeV}, ~~~1/2^-:~ -64.63(2) ~{\rm MeV}\, .\nonumber
\end{equation}
The numbers in parenthesis show the uncertainties of the extrapolations of the energies.
This result is compared with the experiment and with the one from JISP16 in Fig.~\ref{11Be}, which shows that the {\it ab initio} NCSM 
with Daejeon16 successfully reproduces
 the parity inversion in $^{11}$Be.
Note that the JISP16 is unable to reproduce the \mbox{parity} inversion (the current evaluation of the uncertainties
of the JISP16 results is yet preliminary).

For an earlier study of the parity inversion in {\it ab initio} nuclear theory, we refer to Ref.~\cite{Calci:2016dfb}.

\subsection{Deep learning for {\itshape ab initio} nuclear theory}
Recently, we proposed a feed-forward artificial neural network (ANN) method as an
extrapolation tool for {\it ab initio} nuclear theory~\cite{Negoita:2018yok, Negoita:2018kgi}.
Using the {\it ab initio} NCSM 
with Daejeon16 and the feed-forward ANN method, we predicted
the ground-state energy and the ground-state point-proton root-mean-square (rms) radius of $^6$Li.
We observed that our results are nearly converged at $N_{\rm max} = 70$ (ground-state energy) and
${N_{\rm max} = 90}$ (ground-state point-proton rms radius).
Therefore, we concluded that the designed ANNs are sufficient to produce results for these 
two very different observables
utilizing the NCSM results obtained
in small basis spaces
that exhibit the independence of basis space parameters in the limit of extremely large
matrices~\cite{Negoita:2018yok, Negoita:2018kgi}.

Before closing this Section, we refer to Refs.~\cite{Shirokov:2018und, Shirokov:2018nlj}, where
resonance states such as tetraneutron and $^5$He were studied in the framework of
the single-state harmonic oscillator representation of scattering equations
and the {\it ab initio} NCSM 
with Daejeon16 and some other modern $NN$ interactions.

\section{Summary}
In this contribution, we briefly introduced the Daejeon16 $NN$ interaction. We then presented  some of interesting results from the {\it ab initio} NCSM studies 
with
 Daejeon16 and  with some other  $NN$ interactions such as JISP16.  A remarkable result is that the parity inversion in $^{11}$Be is successfully reproduced in our study with Daejeon16.

 We will continue to use Daejeon16 and some other modern $NN$ interactions
 for various {\it ab initio} nuclear studies to be well-prepared for the forthcoming RI facilities.

\section*{Acknowledgments}
 This work is supported in part by the U.S. Department of Energy under
Grants No.~DESC00018223 (SciDAC/NUCLEI) and No.~DE-FG02-87ER40371,
by the Russian Science Foundation
under Grant No.~16-12-10048, by the Rare Isotope Science Project
of Institute for Basic Science funded by Ministry of Science and ICT and National Research Foundation of Korea
(2013M7A1A1075764), and the Ames Laboratory, operated for the U.S. Department of Energy
under Contract No.~DE-AC02-07CH11358 by Iowa State University.


\end{document}